\documentclass[conference]{IEEEtran}
\ifCLASSINFOpdf
\else
\fi

\usepackage{subfigure}
\usepackage[pdftex]{graphicx}
\usepackage{float}
\usepackage{amsmath}
\usepackage{amssymb}
\usepackage{listings}
\usepackage{outlines}
\usepackage{import}
\usepackage[table]{xcolor}
\usepackage[per=slash, decimalsymbol=comma,loctolang={DE:ngerman,UK:english}]{siunitx}
\usepackage{diagbox}

\usepackage[nolist,nohyperlinks]{acronym}

\usepackage{units}

\usepackage{booktabs}

\begin{document}
\IEEEoverridecommandlockouts

%
\title{Resource Allocation for a Wireless Coexistence Management System Based on Reinforcement Learning}
	

\author{\IEEEauthorblockN{Philip S\"{o}ffker, Dimitri Block, Nico Wiebusch, Uwe Meier}
\IEEEauthorblockA{inIT, Institute of Industrial Information Technologies\\
OWL University of Applied Sciences\\
Liebigstra\ss{}e 87\\
32657 Lemgo, Germany\\
Email: \{philip.soeffker, dimitri.block, nico.wiebusch, uwe.meier\}@hs-owl.de }
}


%


\maketitle

\begin{abstract}
In industrial environments an increasing amount of wireless devices are used, which utilize licence-free bands. As a consequence of this mutual interferences of wireless systems might decrease the state of coexistence. 
Therefore, a central coexistence management system is needed, which allocates conflict-free resources to wireless systems. To ensure a conflict-free resource utilization, it is useful to predict the prospective medium utilisation before resources are allocated.
This paper presents a self learning concept, which is based on reinforcement learning.
A simulative evaluation of reinforcement learning agents based on neural networks, called deep Q-networks and double deep Q-networks, was realised for exemplary and practically relevant coexistence scenarios.
The evaluation of the double deep Q-network showed, that a prediction accuracy of at least 98\,\% can be reached in all investigated scenarios.

\end{abstract}


\begin{acronym}[Bash]
\acro{ACI}{adjacent channel interference}
\acro{BER}{bit error rate}
\acro{CTI}{cross-technology interference}
\acro{ISI}{inter-symbol interference}
\acro{LBT}{listen before talk}
\acro{LQI}{link quality indicator}
\acro{PDU}{protocol data unit}
\acro{PLR}{packet loss ratio}
\acro{QoC}{quality-of-coexistence}
\acro{QoS}{quality of service}
\acro{RTSA}{real-time spectrum analyzer}
\acro{SNMP}{simple network management protocol}
\acro{SWT}{standardized wireless technology} \acrodefplural{SWT}{standardized wireless technologies}
\acro{UWB}{ultra wide-band}
\acro{VSG}{vector signal generator}
\acro{WCS}{wireless communication system}
\acro{WN}{wireless network}
\acro{WT}{wireless technology} \acrodefplural{WT}{wireless technologies}

\acro{CSMA/CA}{carrier-sense multiple access with collision avoidance}
\acro{DSSS}{direct-sequence spread spectrum}
\acro{MAC}{medium access control}
\acro{OFDM}{orthogonal frequency-division multiplex}
\acro{PHY}{physical layer}
\acro{PSK}{phase-shift keying}

\acro{BLE}{Bluetooth Low Energy}
\acro{BT}{Bluetooth}
\acro{HART}{highway addressable remote transducer protocol}
\acro{IWLAN}{industrial WLAN}
\acro{LR-WPAN}{low-rate wireless personal area network}
\acro{LTE}{3GPP long-term evolution}
\acro{PNO}{Profibus Nutzerorganisation}
\acro{WLAN}{wireless local area network}
\acro{WSAN-FA}{wireless sensor and actuator network
for factory automation}

\acro{BA}{building automation}
\acro{FA}{factory automation}
\acro{IoT}{internet of things}
\acro{PA}{process automation}

\acro{AWGN}{additive white Gaussian noise}
\acro{CW}{continuous-wave}
\acro{DC}{direct current}
\acro{ESD}{energy spectral density}
\acro{ESD}{energy spectral density}
\acro{FMCW}{frequency-modulated CW}
\acro{ISM}{industrial, scientific and medical}
\acro{LO}{local oscillator}
\acro{MIMO}{multiple-input and multiple-output}
\acro{MR}{medium request}
\acro{PAPR}{peak-to-average-power ratio}
\acro{PDP}{power delay profile}
\acro{PLE}{path-loss exponent}
\acro{PSD}{power spectral density} \acrodefplural{PSD}{power spectral densities}
\acro{RF}{radio frequency}
\acro{RSA}{real time spectrum analyzer}
\acro{SINR}{signal-to-interference-plus-noise ratio}
\acro{SIR}{signal-to-interference ratio}
\acro{SNR}{signal-to-noise ratio}
\acro{USRP}{universal software radio peripheral}
\acro{VSG}{vector signal generator}

\acro{DSP}{digital signal processing}
\acro{HDR}{hardware-defined radio}
\acro{OOT}{out-of-tree}
\acro{SDR}{software-defined radio}

\acro{ACF}{auto-correlation function}
\acro{CCF}{cross-correlation function}
\acro{FZC}{Frank Zadoff Chu}
\acro{MLS}{maximum length sequence}
\acro{PACF}{periodic auto-correlation function}
\acro{PCCF}{periodic cross-correlation function}
\acro{PN}{pseudo-noise}

\acro{TOSM}{through-open-short-match}

\acro{LOS}{line-of-sight}
\acro{NLOS}{non-line-of-sight}
\acro{OLOS}{obstructed line-of-sight}

\acro{FFT}{fast Fourier transform}
\acro{IQ}{in-phase and quadrature}
\acro{STFT}{short-term Fourier transform}

\acro{AGC}{automatic gain control}
\acro{FPGA}{field-programmable gate array}
\acro{LFSR}{linear-feedback shift register}

\acro{CCP}{central coordination point}
\acro{GNSS}{global navigation satellite system}
\acro{PMT}{polymorphic type}
\acro{OS}{operating system}
\acro{SBC}{single-board computer}
\acro{C-RAN}{centralized radio access network}

\acro{CNN}{convolutional neural network}
\acro{DDQN}{double deep Q-network}
\acro{DNN}{deep neural network}
\acro{DQN}{deep Q-network}
\acro{RL}{reinforcement learning}

\acro{NFSC}{neuro-fuzzy signal classifier}
\acro{SDCM}{software-defined coexistence management}
\acro{WIC}{wireless interference classification}

\end{acronym}

%
\IEEEpeerreviewmaketitle

\section{Introduction}\label{sec:introduction}

License-free \ac{RF} bands such as the 2.4-GHz-ISM band are shared between incompatible heterogeneous wireless communication systems.
In industrial environments, typically standardized \acp{WCS} within this band are wide-band high-rate IEEE 802.11 called \ac{WLAN}, narrow-band low-rate IEEE 802.15.4-based WirelessHART and ISA 100.11a, and IEEE 802.15.1-related PNO WSAN-FA and \ac{BT}. 
Additionally, the spectrum band is shared with many proprietary \acp{WT} which target specific application requirements.
Hence, sharing the spectrum may cause interferences between these heterogeneous \acp{WT}.


Therefore, the norm IEC 62657-2 \cite{2013.IEC} for industrial \acp{WCS} recommends an active coexistence management for reliable medium utilization and mitigation of interferences. 
The IEC recommends the use of a (i) manual, (ii) automatic non-cooperative or (iii) automatic cooperative coexistence management.
The first approach is the most inefficient one, due to time-consuming complex configuration effort. 
The automatic approaches (ii) and (iii) enable efficient self-reconfiguration without manual intervention and radio-specific expertise. 
An automatic cooperative coexistence management (iii) requires a control channel, i.e. a logical common communication connection between each coexisting wireless system to enable deterministic medium access. 
In case of a single legacy coexisting wireless system without such connection, the non-cooperative approach (ii) is recommended. 
Non-cooperative coexistence management approaches may also utilize cooperative \acp{WCS} but are able to react on non-cooperative \ac{WCS} which cause independent interferences.
Such coexistence managements require approaches, which mitigate temporary interferences but also predict future medium utilization.
Hence, a self-optimizing resource allocation behavior of the remaining cooperative \acp{WCS} is required.

Additionally, heterogeneity in industrial environments leads to high complexity. 
Hence, coexistence management requires self-learning approaches which models the dynamic heterogeneity of the industrial environment.

\begin{figure}
	\centering
	\includegraphics[width=0.7\linewidth]{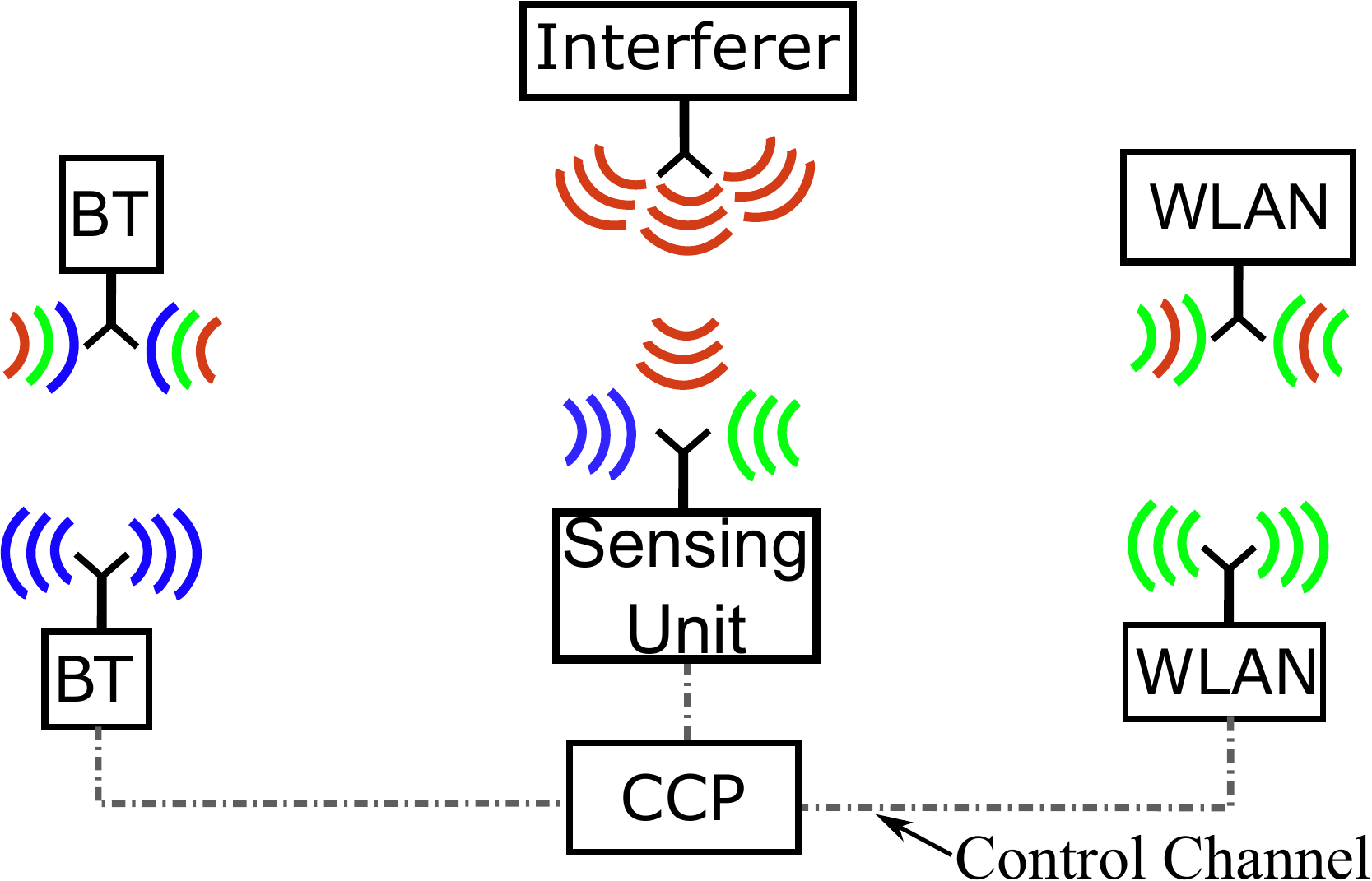}
	\caption{Structure of a central coexistence management with two \acp{WCS} and one interferer}
	\label{fig:CoexMgmtStructure}
	\vspace*{-0.5cm}
\end{figure}

In particular, the actual non-cooperative coexistence managements are able to manage the resource allocation of connected \ac{WCS}.
The control channel can be used for a bidirectional communication of arbitrary \acp{WCS} and a dedicated coexistence management entity, which is called \ac{CCP}. 
So the \ac{CCP} can allocate resources to the different \acp{WCS}. 
This coexistence management principle is shown in Fig. \ref{fig:CoexMgmtStructure} with \ac{BT} and \ac{WLAN} as heterogeneous \acp{WCS}.
There are also \acp{WCS} which are not connected to the \ac{CCP}. 
So the coexistence management cannot control these \acp{WCS}. Therefore, these uncontrollable \acp{WCS} are called interferers.
A sensing unit observes the current \ac{RF} spectral emissions as resource information for the \ac{CCP}. 
The \ac{CCP} uses these current informations to allocate free resources to the \acp{WCS}, without predicting the possible utilization of these allocated resources.





In this paper we propose a resource allocation concept for industrial non-cooperative coexistence management with a \ac{RL} approach.
This approach learns to predict the future medium utilization. 
Additionally, self-optimization improves the resource allocation behavior.


In general, \ac{RL} \cite{2018.SuttonBarto} targets self-optimization and prediction problems of agents which interact with the surrounding environment.
The agents observe the environment, take decisions based on the observations, and are rewarded therefore by the environment.
A \ac{RL} approach enables self-optimization without requirements of certain problem-domain knowledge.

For resource allocation within industrial coexistence management systems, a \ac{RL} agent observes \ac{RF} spectral emissions from utilized as well as from interfering \acp{WCS}.
Then, the agent has to take decisions for resource allocation, which involves for example spectral, temporal and transmission power adjustments for the utilized \acp{WCS}.
The \acp{WCS} apply and evaluate the adjustments based on various quality indicators such as \ac{LQI} and \ac{PLR}.
Based on the quality indicators the \ac{RL} agent get rewarded for its decision.



The following section II presents the related work. Section III will explain the concept of reinforcement learning based resource allocation for a central coexistence management system. 
This leads to section IV, where the presented concept will be simulated for exemplary and practically relevant coexistence scenarios. 
The results of this simulation will be presented in section V. Finally, section VI concludes the paper.

\section{Related Work}\label{sec:RelWork}



\begin{outline}
The requirement for a self-optimizing resource utilization approach based on \ac{RL} was already proposed by Ren et al. \cite{2010.RenDmochowski} in 2010.
They use a special Q-learning algorithm \cite{2018.SuttonBarto} to find and predict non-utilized time intervals called whitespaces within licensed \ac{RF} bands. 
Moreover, the distributed \acp{WCS} do this prediction autonomously.  
Additionally, there is no consultation between the individual distributed \acp{WCS} and no management entity.
So each \ac{WCS} acts opportunistically. 

Liu et al. \cite{2017.LiuYoo} propose a \ac{RL} coexistence management approach for a time-slotted medium access of LTE-U and \ac{WLAN} systems in license-free \ac{RF} bands.
Therefor, they also use a Q-learning algorithm to allocate dynamically free time slots to the LTE-U and \ac{WLAN} systems. 	
They assume, that there are no other \acp{WCS} with different \acp{WT}, which use the \ac{RF} band at the same time. 
So it is an exclusively occupied \ac{RF} band.
Even if there are other \acp{WCS} with different \acp{WT} it is assumed, that these technologies are known and a influence on them is possible.
So it is a cooperative coexistence management, which cannot handle non-cooperative \ac{WCS}.
In \cite{2017.LiuYoo} the individual \acp{WCS} communicate directly with each other to negotiate for time slots. They do not use a central management entity. 

A neural net based \ac{RL} approach is used in \cite{2017.NaparstekCohen} for a distributed medium access.
In that approach one medium utilization solution is trained at a single central unit for all distributed \ac{WCS}. 
The approach is limited to orthogonal resource utilization, which can not be assumed for heterogeneous \acp{WT}.
This single trained solution is transfered to all distributed \acp{WCS} and rarely updated. 
Afterwards every \ac{WCS} uses the trained solution to access the medium independently. 
So there is no consultation among the several distributed \acp{WCS} or between \acp{WCS} and central unit to manage the spectrum access.

All showed approaches act opportunistic and predict only for themselves or their \ac{WT} which resources will be occupied in the future.
Hence, there is no \ac{RL} approach, which proactively predicts the medium utilization for heterogeneous \acp{WCS} in a central non-cooperative coexistence management.

	

		
\end{outline}


\section{Concept}\label{sec:Concept}

The allocation of conflict-free resources is a fundamental part of a coexistence management system. 
However, before a resource allocation can be executed, a prediction of the future medium utilization is required.
Such a central prediction and resource allocation is achieved in this paper with a \ac{RL} approach.
Hence, the coexistence management system can be described by a \ac{RL} structure.
As mentioned before, such a \ac{RL} structure basically consist of two parts.
The first part is the environment.
This environment is a coexistence management environment, e.g. a shop floor.
The second part is the agent.
This agent is the \ac{CCP} of the coexistence management system.

\subsection{Coexistence Management Environment}

Shop floor environments are often equipped with metallic objects like machines. 
These cause reflections, absorptions and dispersions.  
Additionally, the environment contains many \acp{WCS}. 
These \acp{WCS} are usually organized in \acp{WN}.
Thereby, $N$ \acp{WN} use $M$ different \acp{WT}, with $M \leq N$. 
\begin{figure}
	\centering
	\includegraphics[width=0.65\linewidth]{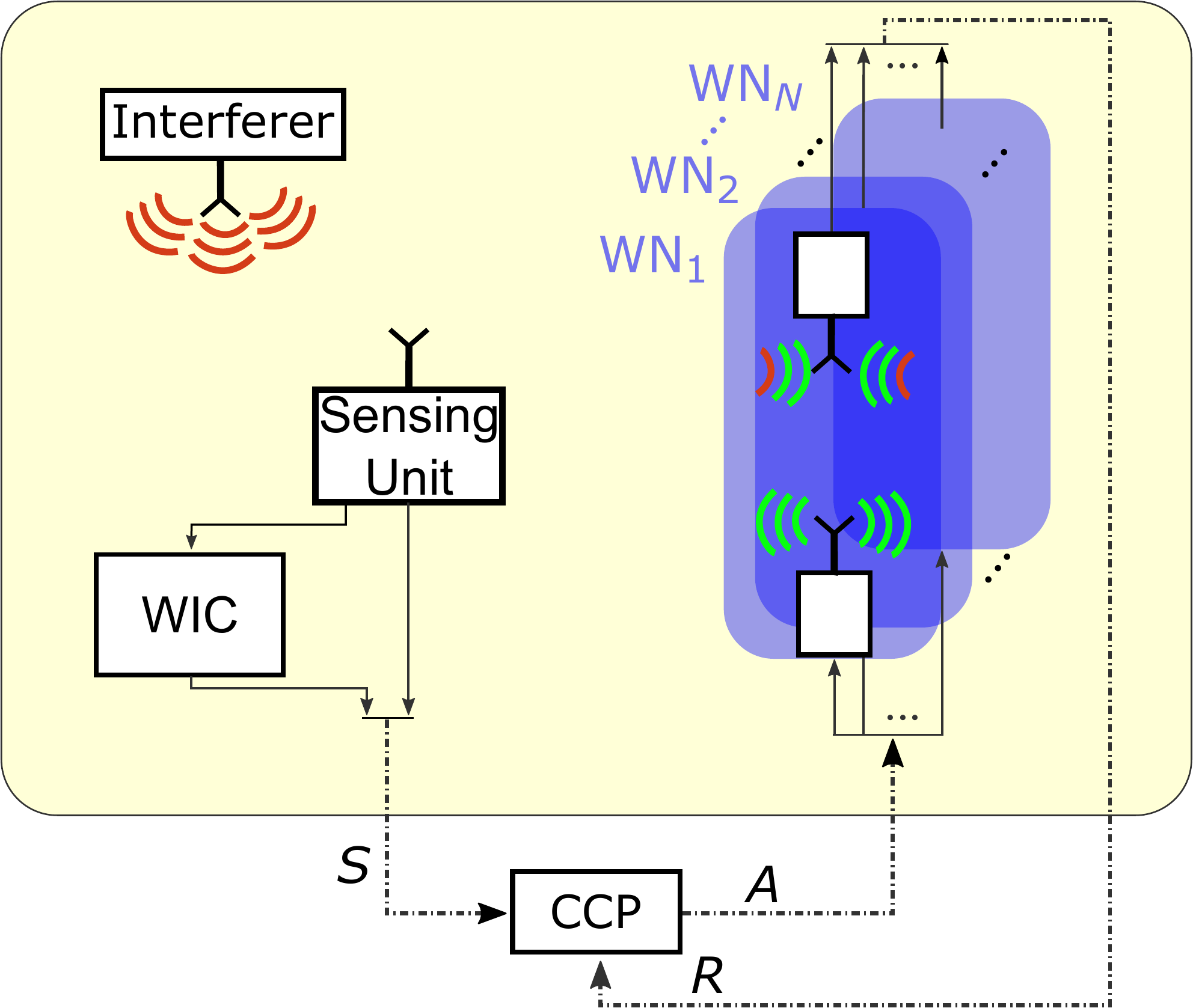}
	\caption{Architecture of the \ac{RL}-based coexistence management environment, and interactions between \ac{CCP} and coexistence management environment}
	\label{fig:CoexEnv}
	\vspace*{-0.5cm}
\end{figure}
Despite the heterogeneity of the \acp{WN}, a control channel is required to link the different \acp{WN} with the \ac{CCP}. 
Such a control channel can be realized, like in \cite{2017.WiebuschBlock}, with Ethernet, whereby \ac{SNMP} is used.
Furthermore, there are emitting interferers in the environment.
Such an environment is illustrated in Fig. \ref{fig:CoexEnv}, with the \ac{CCP}, \acp{WN} and interferers.

The actual state of the coexistence management environment is captured for the \ac{CCP} as observation \textit{\textbf{S}}.
This observation consists of two separate observation elements.
The first element captures the \ac{RF} spectral emissions from \acp{WN} and interferers with the aid of a sensing unit. 
This element consists of discrete-time samples which are captured for a fixed time interval.
Afterwards, these samples are transformed with a \ac{FFT} and the magnitude of the spectrum is computed to $| S_\text{FFT} |$. 
Hence, the magnitude of the spectrum is like a snapshot 
of the coexistence management environment.
The second observation element is an analysis of the captured \ac{RF} spectral emissions for the same fixed time interval.
For example this analysis classifies the captured spectrum on occupied frequency channels of a-priori known \acp{WT}. 
Hence, non-cooperative interfering \acp{WCS} can be classified. 
This is called \ac{WIC}. 
It is possible to classify different \acp{WT} simultaneously, which is helpful in crowded wireless environments.
Such \acp{WIC} are \ac{NFSC} \cite{2016.BlockTows} or \ac{CNN} approaches \cite{2017.SchmidtBlock,2018.GrunauS.BlockD.Meier}.
So the observation can be written as:
\begin{align}
	\label{eq:statesGeneral}
	\textbf{\textit{S}} &= 
	\begin{pmatrix}
		|S_{\text{FFT}}|\\
		S_{\text{WIC}}\\
	\end{pmatrix}
\end{align}

Based on the observation, the agent performs actions \textit{\textbf{A}}. 
For a wireless coexistence management system these actions are resource allocations.
They are allocated by the \ac{CCP}.
Each \ac{WN} gets its own dedicated resource $a$ allocated:
\begin{equation}
\label{eq:actions}
\textbf{\textit{A}} = 
\begin{pmatrix}
a_1	&	a_2	&	a_3	& \ldots	&	a_N
\end{pmatrix}^\text{T}
\end{equation}
Some \acp{WN} use a static channel selection such as one based on \ac{WLAN}. These \acp{WN} get for example an allocation of a single resource, which is a frequency channel.
However, other \acp{WN} use frequency hopping such as one based on \ac{BT}. These \acp{WN} get for example an allocation of multiple frequency channels. 
So a bunch of channels are allocated, whereby the \ac{WN} can select its own specific channel among them.
Thus there are two types of resources, which have to be handled by the \ac{CCP}.

The \acp{WN} use these allocated resources which have to be evaluated with a reward \textit{\textbf{R}}.
The reward is derived from the quality of data transmission on the allocated resources.
This quality of data transmission can be expressed as a metric like the \ac{QoC} parameter \cite{2017.WiebuschSoffker}.
The \ac{QoC} parameter aggregates transmission-related characteristics, such as transmission time, update time and \ac{PLR}.
Each \ac{WN} evaluates the \ac{QoC} for itself with the scalar value $QoC_{\text{WN}_i}$.
If other interfering systems use these allocated resource too, then the value of the \ac{QoC} will decrease.
This feedback is used, to validate the resource allocation of the \ac{CCP} to each \ac{WN} as reward:
\begin{align}
\label{eq:rewards}
\textbf{\textit{R}} =
\begin{pmatrix}
R_1	&	R_2		& R_3	&\ldots	&	R_N
\end{pmatrix}^\text{T} \, \forall \, R_i = QoC_{\text{WN}_i}
\end{align}
Hence, this \ac{QoC}-parameter is a feedback for each resource allocation.

%

\subsection{Central Coordination Point}

The \ac{CCP} is the \ac{RL}-agent.
It learns to allocate conflict free resources to the \acp{WN}. 
However, at the beginning the \ac{CCP} has no a-priori knowledge about the coexistence management environment.
Therefore, the \ac{CCP} has to optimize its resource allocation and learn to predict the occupancy of resources.


The \ac{CCP} interacts with the environment in multiple steps. 
At the beginning of each step the \ac{CCP} checks the initial observation \textit{\textbf{S}}.
Based on this observation, the \ac{CCP} takes decisions for actions \textit{\textbf{A}} for all \acp{WN}.
The \acp{WN} perform their data transmissions on the resources, which were allocated by the actions.
Then each data transmission is individually rewarded with the \ac{QoC} parameter, which is an evaluation of the actions.
Because of these actions the \ac{CCP} needs a new observation for evaluating the new environmental situation.
This new observation is the last part of each step and is also the basis for the next step.
Each step follows this mentioned order.

The \ac{CCP} interacts in two phases with the environment.
The first phase is the training phase, wherein the \ac{CCP} is trained for a particular environment.
The second phase is the operational phase, wherein the \ac{CCP} has completed its training but still has the ability for minor optimizations.
These two phases are divided into episodes. 
Each episode has a predefined number of steps, e.g. 20 steps form an episode.
After each episode the environment can be reset, which are like arbitrary processing time gaps.
Meanwhile the interferer may change their resource, which is a challenge for the \ac{CCP}.
The \ac{CCP} keeps his learned knowledge, despite the reset of the environment.  
Hence, the \ac{CCP} learns a policy to maximize its reward.
Thus it learns when to allocate which conflict-free resource to what \ac{WN}, by predicting the occupation of resources due to interferers. 
This process of autonomously learning the behavior of interferers and the consequent opportunity of a prediction and allocation of free resources, is the advantage of a \ac{RL} based central coexistence management system.

\begin{figure}
	\centering
	\includegraphics[width=0.95\linewidth]{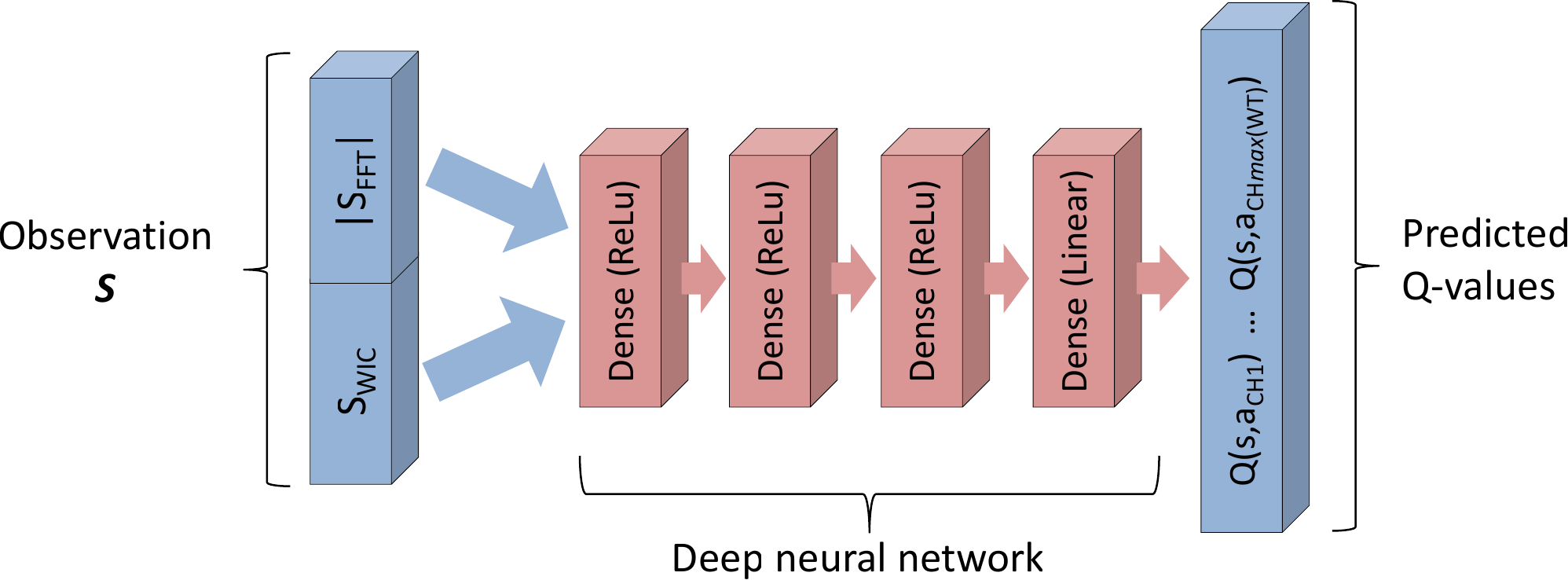}
	\caption{Architecture of a \ac{DQN} or \ac{DDQN} With observation as input and predicted quality of \ac{WT} specific resources as output}
	\label{fig:DqnArchitecture}
	\vspace*{-0.5cm}
\end{figure}

The \ac{CCP} applies learning approaches.
A well known technique of learning is Q-learning \cite{2018.SuttonBarto}.
It uses Q-values, which addresses in this context the predicted quality of \ac{WT} specific resources, a learning rate $\alpha$, which describes how fast newly learned knowledge overrides old knowledge, and a discount factor $\gamma$, which weights the relevance of immediate and future rewards.
So the \ac{CCP} learns by comparing predicted Q-values with the reward of the actual executed actions.
This Q-learning can be extended with neural networks.
Such an extension is helpful for large observation spaces and is called \ac{DQN} \cite{2015.MnihKavukcuoglu}.
These large observation spaces are also given with a large \ac{FFT} length.
Another problem of wireless environments is, that they are noisy.
For such cases van Hasselt \cite{2010.HadoV.Hasselt} proposed the double Q-learning approach, which uses two independent Q-functions.
So even if one Q-function is biased it is not correlated to the other one, which reduces the prediction error.
This double Q-learning was also combined with neural networks, which leads to \acp{DDQN} \cite{2016.vanHasseltGuez}.
The \ac{CCP} either uses a \ac{DQN} or a \ac{DDQN}. Each of them uses a \ac{DNN}
, as pictured in Fig. \ref{fig:DqnArchitecture}. 
Additionally experience replay is used for both \ac{CCP} types and a target network \cite{2016.vanHasseltGuez} is used at the \ac{DDQN}.


\section{Simulation \& Results}\label{sec:Simulation}




The simlation can be divided into coexistence management environment and \ac{CCP}.
For the environment the signal processing framework GNU Radio\footnote{\texttt{gnuradio.org} version 3.7.11 (27 Feb., 2017)} is used. 
It enables the simulation of multiple \acp{WCS}, interferers and noise.
For the proof of concept a lean realization is addressed which could be scaled later on.
Therefore, the environment only processes synchronous streams without asynchronous events.

The lean realization is limited to a single \ac{WN}, i.e. $M=1$
This \ac{WN} consist of two \acp{WCS} which utilize unidirectional communications. 
The \ac{WN} uses a static channel selection with four possible channels.
These four channels are the action space.
Then, the transmitting \ac{WCS} applies a \ac{PSK} modulation.
Afterwards the receiving \ac{WCS} demodulates the transmitted data.
Additionally for simplicity, it calculates the reward from the \ac{BER} for each step with $R=1-\text{\ac{BER}}$.

The data transmission is efficiently disturbed by an interferer.
For a worst-case disturbance, this interferer inverses the \ac{PSK} modulation of the \ac{WN}. 
Hence, the \acp{PSD} of the interferer and \ac{WN} are almost indistinguishable.  
This interferer will be a problem for the \ac{WN}, if both use the same frequency channel.
For the interference two coexistence scenarios are used: 
(i) static interferer, where the interferer occupies a channel for the duration of an episode like \ac{WLAN}, and (ii) sequential hopping interferer, where the interferer sequentially changes the channel after each step like WirelessHART.
In both scenarios the interferer chooses for each episode the initial channel randomly.

The sensing unit capture the complete band of all four channels for the duration of each step.
It contains 1024 I/Q-samples, which is used for the generic observation element $| S_\text{FFT} |$.  
The additional specific \ac{WIC} observation element is omitted, because of the lean realization.

For the simulation of the \ac{DQN} and \ac{DDQN} \ac{CCP} OpenAI Gym\footnote{\texttt{gym.openai.com} version 0.7.4 (5th Mar., 2017)} is used.
Both networks use a \ac{DNN} with four dense layers with the output size of 256, 64, 32 and 4, respectively.
Additionally, they use experience replay and the \ac{DDQN} uses a target network.
Further hyperparameters of the networks are listed in Table \ref{tab:Hyperparameter}.

Each simulation experiment consist of 250 episodes and is repeated 15 times.
The first 100 episodes are the training phase.
It takes place with the help of an $\mathcal{E}$-greedy exploration approach.
The remaining 150 episodes are the operational phase.
Each episode is separated into 20 steps.
So, random frequency channel choice results in an average accumulated reward of 15.
\begin{table}[htbp]
	\centering
	\caption{Hyperparameters of the \ac{CCP}}	
	\label{tab:Hyperparameter}
	\begin{tabular}{l|c}
		Hyperparameters							& Values	 \\ 
		\hline
		Learning rate $\alpha$ 					& 0,0001 \\ 
		Discount factor $\gamma$				& 0,96  \\ 
		Minibatch size							& 32 \\ 
		Initial exploration	(training phase)	& 1\\
		Final exploration (operational phase)	& 0,01\\
		Update frequency of target network		& 20 episodes\\		
	\end{tabular}
\end{table}


\begin{figure}
	\centering
	\subfigure[Scenario (i) with \ac{DQN} \ac{CCP}]
	{
		\includegraphics[width=0.8\linewidth]{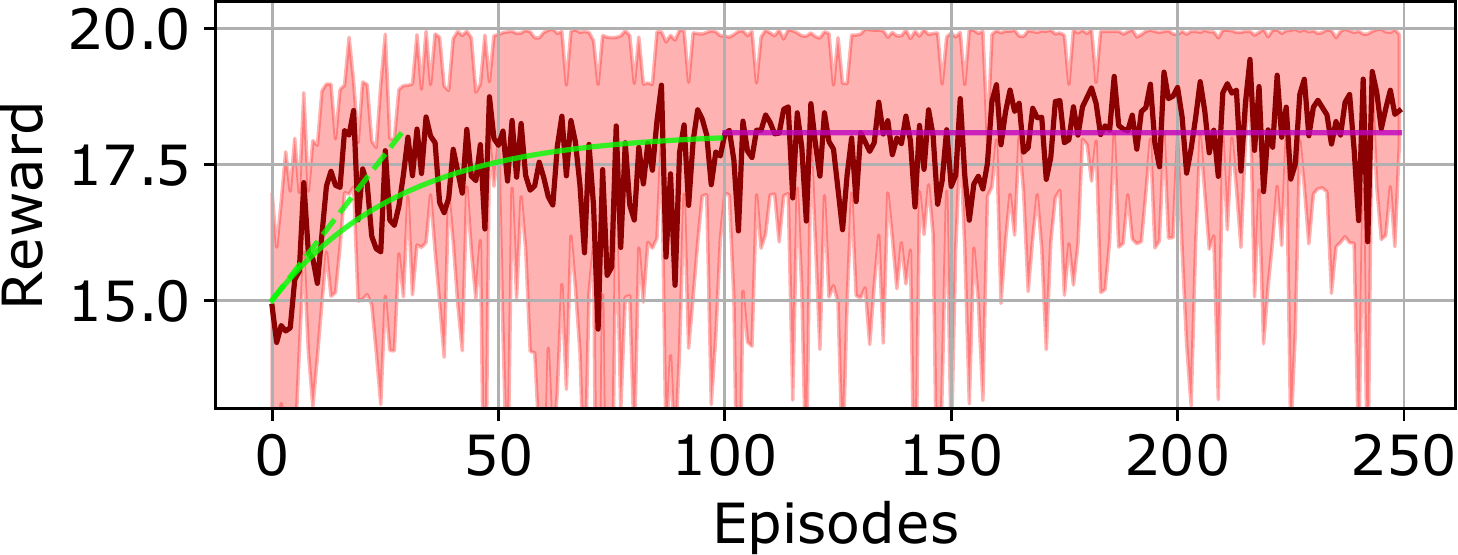}
		\label{fig:scen_i_DQN}
	}
	\subfigure[Scenario (i) with \ac{DDQN} \ac{CCP}]
	{
		\includegraphics[width=0.8\linewidth]{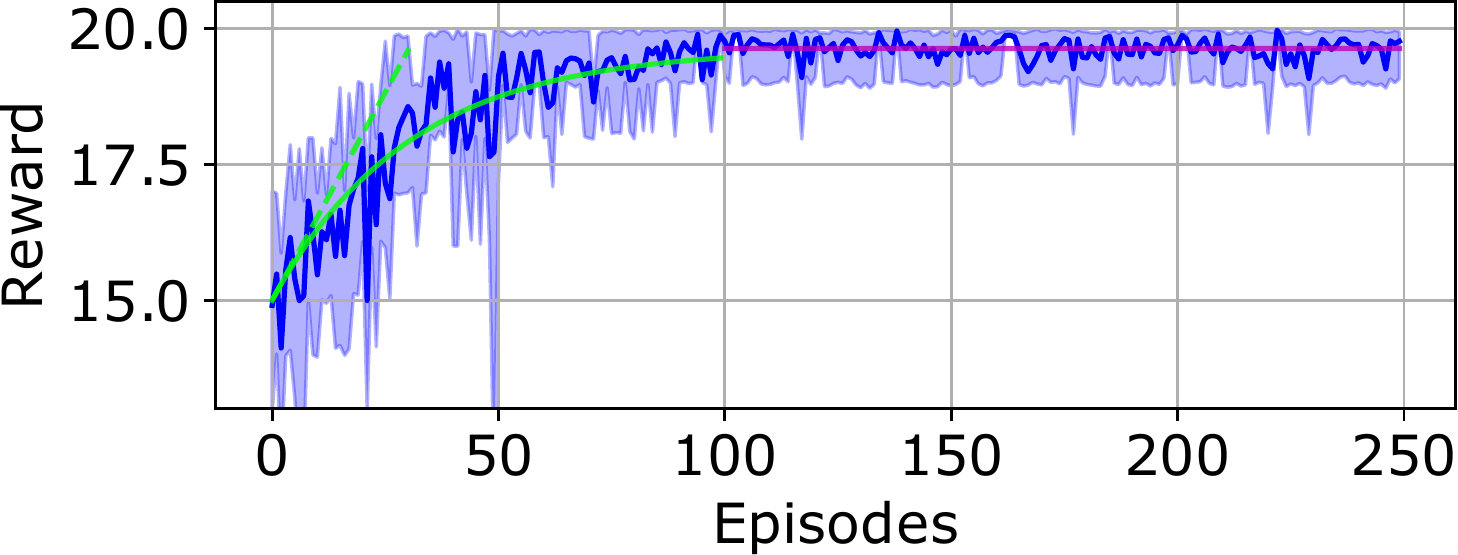}
		\label{fig:scen_i_DDQN}
	}
	\subfigure[Scenario (ii) with \ac{DQN} \ac{CCP}]
	{
		\includegraphics[width=0.8\linewidth]{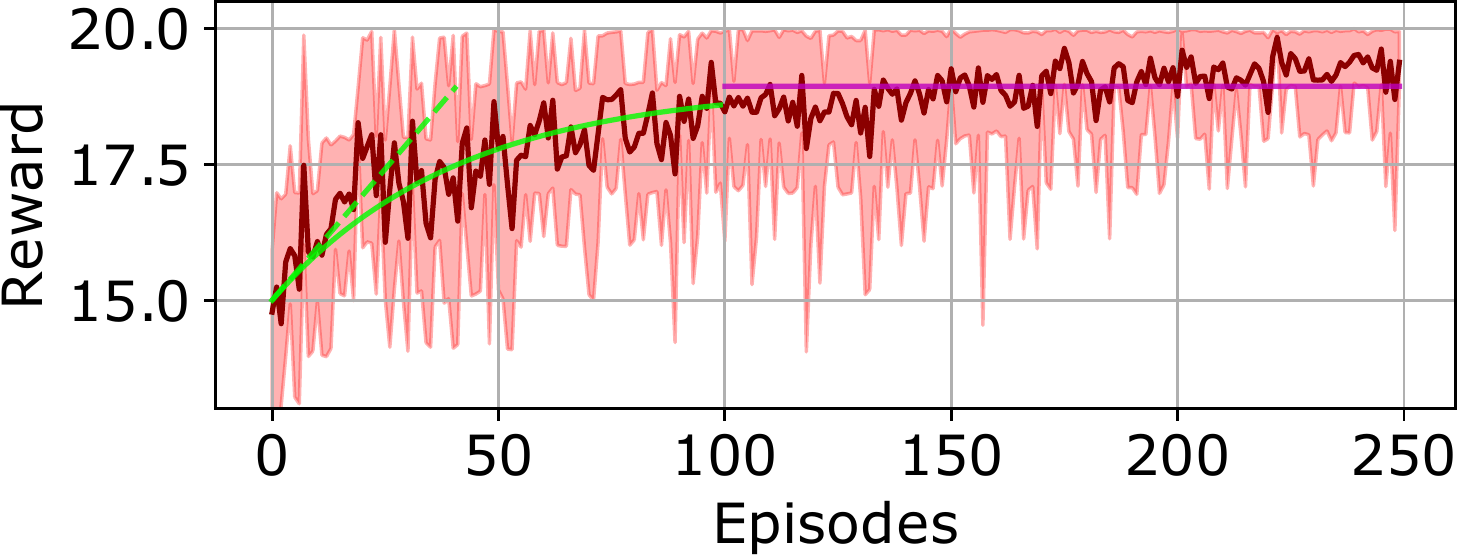}
		\label{fig:scen_ii_DQN}
	}
	\subfigure[Scenario (ii) with \ac{DDQN} \ac{CCP}]
	{
		\includegraphics[width=0.8\linewidth]{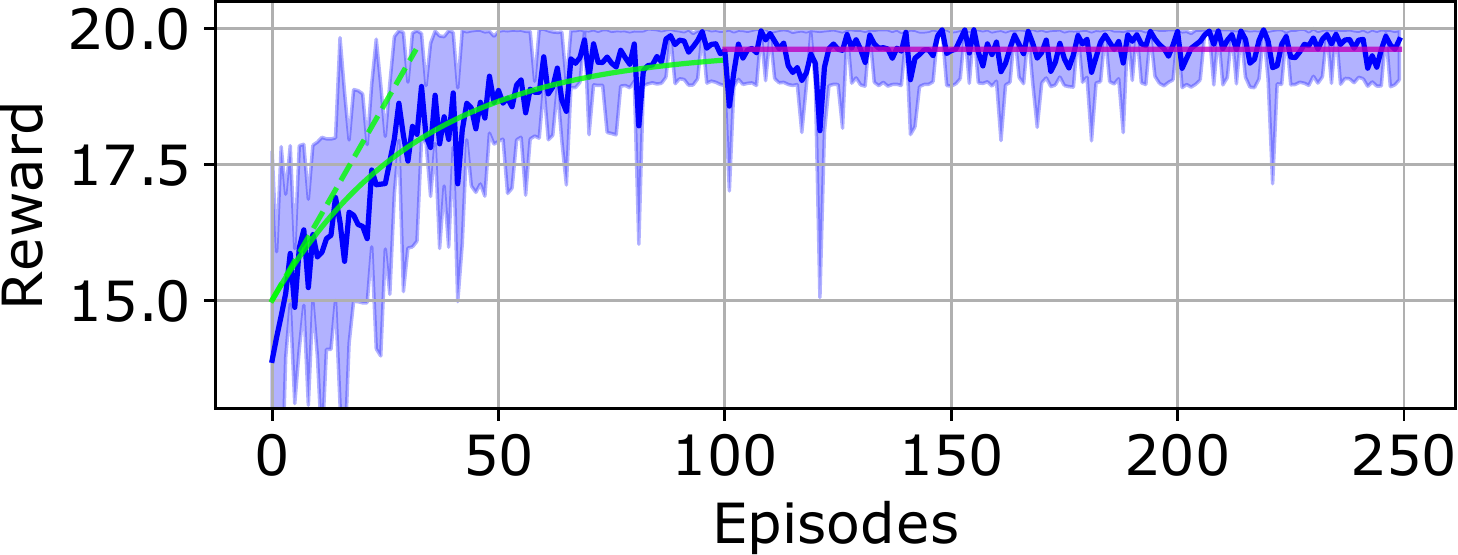}
		\label{fig:scen_ii_DDQN}
	}
	\caption{Simulation results of \ac{DQN} and \ac{DDQN} based \ac{CCP} of the scenario (i) and (ii); the solid lines are the mean reward, the shaded areas represent the values between the 10\% and 90\% percentile}
	\label{fig:Simulation_results}
\end{figure}

The results of the simulation are shown in Fig. \ref{fig:Simulation_results}.
For a better comparison of the operational phase a mean reward is estimated.
This mean reward is shown in Fig. \ref{fig:Simulation_results} as purple colored line.
Thereby, the \ac{DDQN} \ac{CCP} reaches in all scenarios a mean reward of at least 19,6 which corresponds to a prediction accuracy of circa 98\%, see Fig. \ref{fig:scen_i_DDQN} and \ref{fig:scen_ii_DDQN}.
That implies, nearly all 20 data transmissions in each episode are received correctly, because of an appropriate resource allocation by the \ac{CCP}.
The \ac{DQN} \ac{CCP} in the first scenario has a mean reward of 18,08 which is not noteworthy in comparison to the other \ac{DQN} and \acp{DDQN}.

For the training phase the exponential learning behavior is estimated as shown in Fig. \ref{fig:Simulation_results} with the green solid line.
The rise time of the exponential behavior is shown in the same figure as green dashed line.
The trainings phase of the \ac{DQN} in scenario (ii) has a rise time of 
In the second scenario the \ac{DDQN} \ac{CCP} results in a rise time of 31.8 episodes. It is 8.9 episodes faster than the \ac{DQN}.
Within the first scenario, the \ac{DDQN} also shows a comparable rise time of 30.3 episodes.


\section{Conclusion}\label{sec:Conclusion}

In this paper we suggested a new concept for a self-optimizing central coexistence management system.
This concept used a deep \ac{RL} approach, which learns to predict future medium utilizations and concludes this knowledge to allocate unoccupied frequency channels to \acp{WN}.
The \ac{RL}-based concept was evaluated with a simulation as proof of concept.
This simulation considered a wireless environment with practically relevant coexistence scenarios.
Additionally, the \ac{CCP} applied \ac{DQN} and \ac{DDQN} based \ac{RL}-agents.
The \ac{DDQN} shows a faster learning process during training and a higher prediction accuracy during operational phase in both scenarios.
Hence, the advantage of such a \ac{RL}-based central coexistence management system is the autonomously learning of interferer behaviors, without any a-priori knowledge about the wireless environment, and the consequential prediction and allocation of unoccupied resources.

In the future \aclp{SDR} have to be integrated to interact with real wireless  environments.



\section*{Acknowledgements}
Parts of this research were funded by KoMe (IGF 18350 BG/3 over DFAM) and HiFlecs (16KIS0266 over BMBF).

\bibliographystyle{IEEEtran}
\bibliography{biblio}

\begin{thebibliography}{10}
\providecommand{\url}[1]{#1}
\csname url@samestyle\endcsname
\providecommand{\newblock}{\relax}
\providecommand{\bibinfo}[2]{#2}
\providecommand{\BIBentrySTDinterwordspacing}{\spaceskip=0pt\relax}
\providecommand{\BIBentryALTinterwordstretchfactor}{4}
\providecommand{\BIBentryALTinterwordspacing}{\spaceskip=\fontdimen2\font plus
\BIBentryALTinterwordstretchfactor\fontdimen3\font minus
  \fontdimen4\font\relax}
\providecommand{\BIBforeignlanguage}[2]{{%
\expandafter\ifx\csname l@#1\endcsname\relax
\typeout{** WARNING: IEEEtran.bst: No hyphenation pattern has been}%
\typeout{** loaded for the language `#1'. Using the pattern for}%
\typeout{** the default language instead.}%
\else
\language=\csname l@#1\endcsname
\fi
#2}}
\providecommand{\BIBdecl}{\relax}
\BIBdecl

\bibitem{2013.IEC}
IEC, ``Industrial communication networks{\~{}}-- wireless communication
  networks{\~{}}-- part 2: Coexistence management,'' 2013.

\bibitem{2018.SuttonBarto}
\BIBentryALTinterwordspacing
R.~Sutton and A.~Barto, ``Reinforcement learning: An introduction: Second
  edition,'' 24.03.2018. [Online]. Available:
  \url{http://incompleteideas.net/book/the-book-2nd.html}
\BIBentrySTDinterwordspacing

\bibitem{2010.RenDmochowski}
Y.~Ren, P.~Dmochowski, and P.~Komisarczuk, ``Analysis and implementation of
  reinforcement learning on a gnu radio cognitive radio platform,'' in
  \emph{Cognitive Radio Oriented Wireless Networks {\&} Communications
  (CROWNCOM), 2010 Proceedings of the Fifth International Conference on}, 2010.

\bibitem{2017.LiuYoo}
Y.-Y. Liu and S.-J. Yoo, ``Dynamic resource allocation using reinforcement
  learning for lte-u and wifi in the unlicensed spectrum,'' in \emph{2017 Ninth
  International Conference on Ubiquitous and Future Networks (ICUFN)}, 2017,
  pp. 471--475.

\bibitem{2017.NaparstekCohen}
O.~Naparstek and K.~Cohen, ``Deep multi-user reinforcement learning for dynamic
  spectrum access in multichannel wireless networks,'' in \emph{GLOBECOM 2017 -
  2017 IEEE Global Communications Conference}, 2017, pp. 1--7.

\bibitem{2017.WiebuschBlock}
N.~Wiebusch, D.~Block, and U.~Meier, ``A centralized cooperative snmp-based
  coexistence management approach for industrial wireless systems,'' in
  \emph{2017 IEEE 13th International Workshop on Factory Communication Systems
  (WFCS)}, 2017, pp. 1--4.

\bibitem{2016.BlockTows}
D.~Block, D.~Tows, and U.~Meier, ``Implementation of efficient real-time
  industrial wireless interference identification algorithms with fuzzified
  neural networks,'' in \emph{2016 24th European Signal Processing Conference
  (EUSIPCO)}, 2016, pp. 1738--1742.

\bibitem{2017.SchmidtBlock}
M.~Schmidt, D.~Block, and U.~Meier, ``Wireless interference identification with
  convolutional neural networks,'' in \emph{2017 IEEE 15th International
  Conference on Industrial Informatics (INDIN)}, 2017, pp. 180--185.

\bibitem{2018.GrunauS.BlockD.Meier}
{S. Grunau}, {D. Block}, and {U. Meier}, ``Multi-label wireless interference
  identification with convolutional neural networks,'' \emph{2018 IEEE 16th
  International Conference 2018 in Press}, 2018.

\bibitem{2017.WiebuschSoffker}
N.~Wiebusch, P.~Soffker, D.~Block, and U.~Meier, ``A multidimensional resource
  allocation concept for wireless coexistence management,'' in \emph{2017 22nd
  IEEE International Conference on Emerging Technologies and Factory
  Automation}, 2017, pp. 1--4.

\bibitem{2015.MnihKavukcuoglu}
V.~Mnih, K.~Kavukcuoglu, D.~Silver, A.~A. Rusu, J.~Veness, M.~G. Bellemare,
  A.~Graves, M.~Riedmiller, A.~K. Fidjeland, G.~Ostrovski, S.~Petersen,
  C.~Beattie, A.~Sadik, I.~Antonoglou, H.~King, D.~Kumaran, D.~Wierstra,
  S.~Legg, and D.~Hassabis, ``Human-level control through deep reinforcement
  learning,'' \emph{Nature}, vol. 518, no. 7540, pp. 529--533, 2015.

\bibitem{2010.HadoV.Hasselt}
H.~{Hado V.}, ``Double q-learning,'' in \emph{Advances in Neural Information
  Processing Systems 23}, {J. D. Lafferty}, {C. K. I. Williams}, {J.
  Shawe-Taylor}, {R. S. Zemel}, and {A. Culotta}, Eds.\hskip 1em plus 0.5em
  minus 0.4em\relax {Curran Associates, Inc}, 2010, pp. 2613--2621.

\bibitem{2016.vanHasseltGuez}
H.~{van Hasselt}, A.~Guez, and D.~Silver, ``Deep reinforcement learning with
  double q-learning,'' in \emph{2016 Proceedings of the Thirtieth AAAI}, 2016,
  pp. 2094--2100.

\end{thebibliography}

\end{document}